\documentclass[10pt,twocolumn,showpacs,showkeys,preprintnumbers,amssymb,aps,superscriptaddress]{revtex4-2}

\usepackage{amsmath, amssymb}
\usepackage{graphicx}
\usepackage{color,array}
\usepackage[colorlinks={true}]{hyperref}
\hypersetup{colorlinks=true,linkcolor=red,citecolor=blue,urlcolor=blue}
\usepackage{comment}
\usepackage{xcolor}
\usepackage{color,array}
\usepackage{bm}
\usepackage{subfigure}
\usepackage{amsmath}
\usepackage{lipsum,appendix}
\usepackage{orcidlink}
\usepackage{pstricks}
\usepackage{booktabs}
\usepackage{amsmath}
\usepackage[margin=1in]{geometry}

\begin{document}

\title{Katsura-Nagaosa-Balatsky magnetoelectricity in molecular magnets: \\Bipartite entanglement transfer by means of rotating electric field}

\author{Zhirayr~Adamyan~\!\!\orcidlink{0000-0001-5937-2675}}
\affiliation{Laboratory of Theoretical Physics,
          Yerevan State University,
         1 Alex Manoogian Str., 0025 Yerevan, Armenia}
 \affiliation{CANDLE, Synchrotron Research Institute, 31 Acharyan Str., 0040 Yerevan, Armenia}

 \author{Ani~Chobanyan~\!\!\orcidlink{0009-0008-8852-4277}}
 \affiliation{Philip Morris Armenia LLC - R\&D Research Center, 105 Teryan, 0009 Yerevan, Armenia}

\author{Vadim~Ohanyan~\!\!\orcidlink{0000-0002-7810-7321}}
 \affiliation{Laboratory of Theoretical Physics,
          Yerevan State University,
         1 Alex Manoogian Str., 0025 Yerevan, Armenia}
 \affiliation{CANDLE, Synchrotron Research Institute, 31 Acharyan Str., 0040 Yerevan, Armenia}

\author{Azadeh~Ghannadan~\!\!\orcidlink{0009-0005-1448-0919}}
\affiliation{Institute of Physics, Slovak Academy of Sciences, D\'{u}bravsk\'{a} cesta 9, 84511 Bratislava, Slovakia}

\author{Jozef~Stre{\v c}ka~\!\!\orcidlink{0000-0003-1667-6841}}
 \affiliation{Department of Theoretical Physics and Astrophysics, Faculty of Science of P. J. \v{S}af{\'a}rik University, Park Angelinum 9, 040 01 Ko\v{s}ice, Slovak Republic}

\author{Saeed~Haddadi~\!\!\orcidlink{0000-0002-1596-0763}}
\address{School of Particles and Accelerators, Institute for Research in Fundamental Sciences (IPM), P.O. Box 19395-5531, Tehran, Iran}

\author{Hamid Arian Zad~\!\!\orcidlink{0000-0002-1348-1777}} \affiliation{Department of Theoretical Physics and Astrophysics, Faculty of Science of P. J. \v{S}af{\'a}rik University, Park Angelinum 9, 040 01 Ko\v{s}ice, Slovak Republic}


\begin{abstract}
We investigate quantum entanglement in a spin-1/2 Heisenberg trimer with spin-induced electric polarization described by the Katsura–Nagaosa–Balatsky (KNB) mechanism in the presence of external magnetic and electric fields. The electric field is assumed to lie in the plane of the magnetic ions, allowing its strength and orientation to be tuned independently. We analyze both bipartite and tripartite entanglement and demonstrate that the spin–electric-field coupling provides an efficient mechanism for controlling quantum correlations within the molecular nanomagnet. Depending on the electric-field parameters, the bipartite entanglement can be significantly enhanced or suppressed, while the multipartite entanglement exhibits a rich dependence on the microscopic spin–electric coupling. Most notably, we demonstrate that a rotating in-plane electric field of constant magnitude induces a controllable transfer of bipartite entanglement between different spin pairs. In the symmetric case of homogeneous exchange interactions and uniform KNB coupling, this transfer is found to be nearly ideal, with the bipartite negativity approaching its theoretical maximum for one spin pair while simultaneously vanishing for the remaining pairs. We show that the efficiency of the transfer can be tailored through the exchange interactions, bond geometry, and nonuniform spin–electric coupling. These results establish molecular nanomagnets with KNB spin–electric coupling as a promising platform for the electrical manipulation, steering, and localization of quantum entanglement at the molecular scale.

\end{abstract}


\maketitle

\section{Introduction}
Molecular magnets~\cite{MM_book} play an important role in the current quantum technologies, particularly in the setups of quantum computers and quantum algorithms. As material carriers of qubits, molecular magnets possess a series of advantages: a wide range of tailorable magneto-chemical properties, a long coherence time, nanoscale size, potential for room-temperature operation, and accessibility of schemes with higher-dimensional quantum bits (qudits), etc. The relevance and advantages of molecular nanomagnets have been widely exploited in quantum information technologies and state-of-the-art experiments~\cite{los01, tro11, ste08, ses15}.
As a class of magnetic materials, molecular magnets are often referred to as $0$-dimensional because the underlying lattice models that describe their magnetic properties are just few-body systems. Magnetic molecules are usually composed of a few $d$- and/or $f$-element ions coordinated with non-magnetic ligands, which are usually involved in super-exchange interactions between these metal ions. In molecular crystals, however, the exchange interactions between metal centers belonging to different molecules are negligible. Thus, often a major part of the magneto-thermal properties of molecular magnets can be captured by relatively simple many-body Hamiltonians that describe the exchange interactions between a few magnetic centers of a single molecule. 
Recently, various models of molecular magnets and spin clusters have attracted considerable theoretical attention as platforms for quantum entanglement engineering. Entanglement phenomena have been traditionally studied in frustrated Heisenberg lattices, chains, and  clusters~\cite{ana11,abg11,ana12,szal23}, as well as in simplified Ising--Heisenberg chains developed for specific coordination polymers~\cite{str14,car18,sou19,sou20,str20,gal21,gal21b,gal22,zhe22}. The effects of magnetic fields, anisotropy, and heterogeneous $g$-factors on bipartite entanglement were analyzed in several Heisenberg spin dimers and trimers~\cite{ada20,cen20,eki20,kar20,ada24,ada24a}. Furthermore, growing interest has been directed toward multipartite entanglement in mixed-spin Heisenberg trimers, tetramers, pentamers, and more complex molecular nanomagnets~\cite{ada24b,gha25,zad25b,var25a,var25b,yuripra}, including their potential applications in quantum teleportation, coherence control, and quantum information processing~\cite{ben22,zad22,gha25,zad25a,asadali}.

The main focus of these theoretical studies is to determine the physical conditions that provide optimal schemes for manipulating and enhancing bipartite and multipartite entanglement in spin clusters. For instance, one of the most effective mechanisms of enhancement of the entanglement properties is non-conserving magnetization due to non-uniform $g$-factors~\cite{oha15,tor16,tor18,var19,kro20, bre20, var19,pan20,jap21, bar23}. Our previous research demonstrated the possibility of a substantial increase in entanglement between some pairs of spins in molecular magnets when the $g$ factor of different spins is nonuniform~\cite{ada20, ada24, ada24a, ada24b}. In this paper, we investigate another physical mechanism that allows enhancement, effective control and transfer of bipartite entanglement between pairs of spins in a spin-$1/2$ Heisenberg model of a trinuclear molecular magnet with the triangular magnetic structure, magneto-electric coupling between spins and dielectric polarization given by the Katsura-Nagaosa-Balatsky (KNB) mechanism~\cite{KNB1, KNB2}. The KNB mechanism supposes an appearance of the dielectric polarization into the lattice bond by the redistribution of the electron densities between two magnetic centers from the edges of the bond in the case of a non-collinear spin configuration on them. The corresponding expression for KNB polarization is also lattice geometry sensitive:
\begin{eqnarray}\label{eq:KNB}
\mathbf{P}_{ij}=\gamma_{ij} \mathbf{e}_{ij}\times\mathbf{S}_i\times\mathbf{S}_j,
\end{eqnarray}
where $ \mathbf{e}_{ij}$ is a unit vector from site $i$ to site $j$ and $\mathbf{S}_{i(j)}$ are the corresponding spin operators. The bond-dependent constant $\gamma_{ij}$ reflects the local quantum-chemical environment features of each bond~\cite{KNB1}. For materials with a uniform structure of the bonds between magnetic ions, the corresponding constant is also uniform. Various models of one-dimensional~\cite{bro13, thakur18, XYZ, oha20, mench15, bar21, sznajd18, sznajd19, baran18, sc_KNB, JJJ, oles, stre20, bre2, dis} as well as two-dimensional \cite{esa, bre1} magnetoelectrics with KNB mechanism have been considered recently. The focus of previous studies was the magnetoelectric response by itself. In the present paper, we propose another possible application of the KNB mechanism, considering it as an efficient tool for enhancing and manipulating quantum entanglement in spin models. Using the simplest model containing minimal set of necessary ingredients, a spin-1/2 Heisenberg model of triangular magnetic molecule, we demonstrate that specific scheme of magnetoelectric effect (MEE) when the sample rotates in a homogeneous in-plane electric field of constant magnitude. The so-called rotating MEE (RMEE)~\cite{cen19, cen21} can be used to perform a controlled transfer of the bipartite entanglement between the spin pairs in the model. This simple setup can be easily performed by means of rotating the sample inside the field of flat capacitor, for example. Despite its simplicity, the scheme allows the practical realization of the rather important process of bipartite entanglement transfer between nearest-neighbor spin pairs in a magnetic molecule.

The paper is organized as follows. In Sec.~\ref{sec2}, we formulate the model and present its exact eigenstates and spectrum.  Section~\ref{sec3} is devoted to quantifying entanglement using negativity. In Sec.~\ref{sec4}, we analyze various aspects of the negativity behavior, particularly demonstrating how an electric field in a rotating setup can be used to transfer entanglement between the pair of the spins. The paper ends with a conclusion in Sec.~\ref{sec5}.

\section{The Model and its exact solution}
\label{sec2}

\begin{figure}[t]
    \begin{center}
      \includegraphics[width=75.5mm]{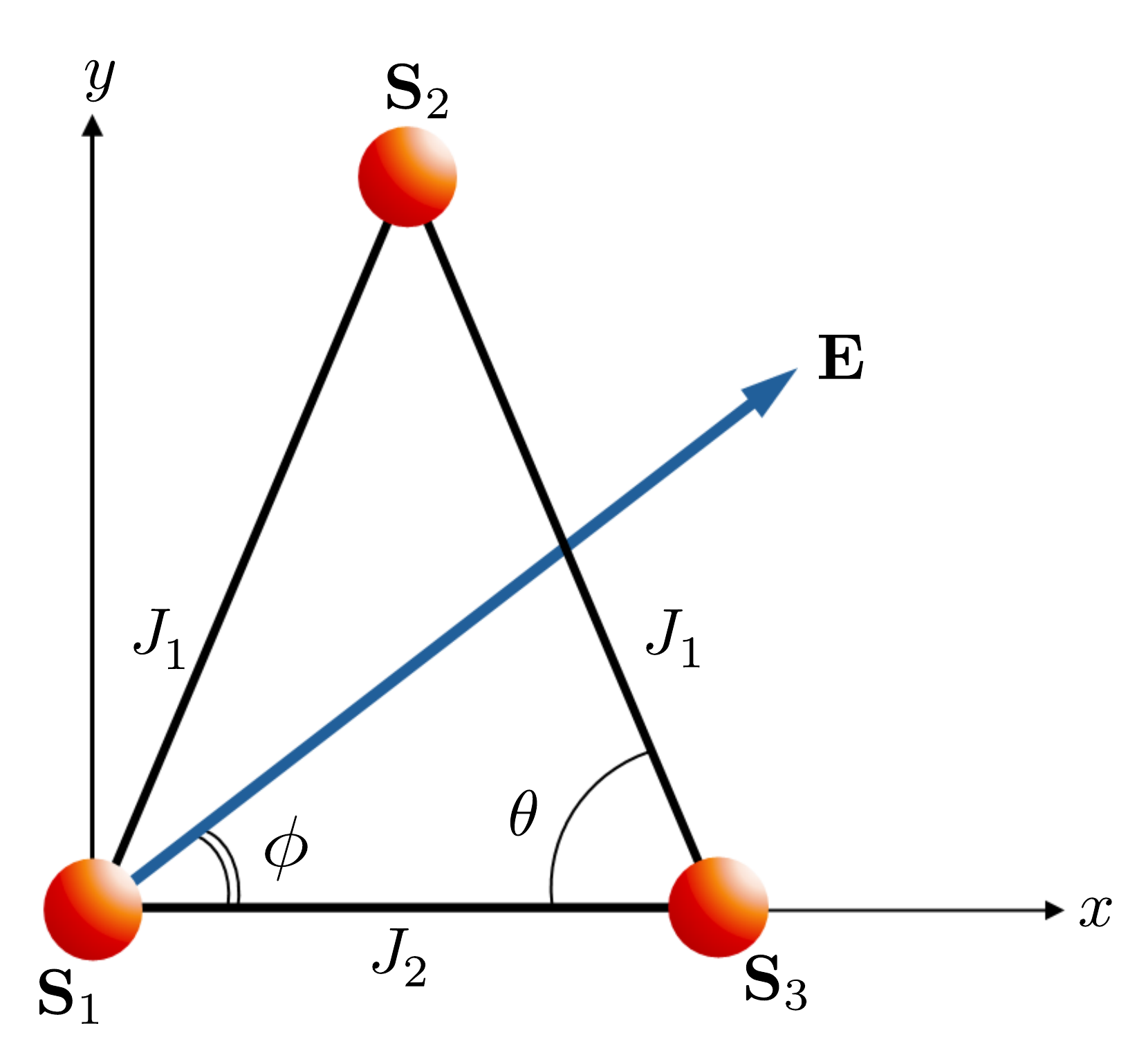}
      \caption{Schematic illustration of the three-spin molecular cluster with an isosceles triangular geometry. The localized $S=1/2$ spins interact via two antiferromagnetic exchange couplings, $J_1$ and $J_2$, and are subjected to an external electric field $\mathbf{E}$ lying in the plane of the triangle. The geometry of the molecule is characterized by the bond angle $\theta$, while the electric-field orientation is specified by the angle $\phi$. Within the KNB mechanism, the spin–electric-field coupling is mediated through bond-dependent DM interactions, whose magnitude and direction are determined by both the molecular geometry and the electric-field orientation. Consequently, the parameters $\theta$ and $\phi$ provide two independent means of controlling the effective spin–electric coupling and, therefore, the magnetic and quantum-entanglement properties of the molecular nanomagnet.
      }\label{fig:1}
    \end{center}
  \end{figure}

We consider a spin-1/2 Heisenberg model for a three-spin cluster, which captures a magnetic structure of a trinuclear molecule consisting of three identical spin-1/2 magnetic ions. The spin cluster is assumed to have the shape of an isosceles triangle with an angle $\theta$ at the base. The geometric features of the cluster influence the exchange Hamiltonian, which includes two symmetric exchange constants: $J_2$, representing the interaction between the spins on the base, and $J_1$, representing the interaction between the apex spin and the two spins forming the base. Another key aspect of the model is spin-induced ferroelectricity, described by the KNB mechanism \cite{KNB1, KNB2}, which introduces additional terms to the spin Hamiltonian when an external in-plane electric field is applied. Figure \ref{fig:1} demonstrates the geometry of the problem and the notation. Consequently, the spin Hamiltonian of the model takes the following form:
\begin{align}\label{Ham}
\mathcal{H}=& J_1\left(\mathbf{S}_1\mathbf{S}_2+\mathbf{S}_2\mathbf{S}_3\right)+J_2 \mathbf{S}_1\mathbf{S}_3\nonumber\\
&-B\left(S_1^z+S_2^z+S_3^z\right)-\mathbf{E} \mathbf{P},
\end{align}
where $B$ is a uniform constant magnetic field directed along the $z$ axis, and the term ``$-\mathbf{E}\mathbf{P}$"  describes the interaction between the KNB polarization of the system and the external uniform constant electric field that is supposed to lie in the plane of the triangle, which in its turn is the $(x, y)$-plane:
\begin{eqnarray}\label{Evec}
\mathbf{E} = \left(E \cos \phi,~E \sin \phi,~0\right).
\end{eqnarray}
With this configuration of fields and the above geometry of the system, the interaction between KNB polarization and the external electric field is given by the effective Dzyaloshinsky-Moriya (DM) interaction \cite{baran18, sc_KNB, JJJ, dis}:
\begin{align}\label{eq: KNB_pol}
-\mathbf{E}\mathbf{P} &=\sum_{i<j} D_{ij}\left(S_i^x S_j^y - S_i^y S_j^x\right), \end{align}
with
\begin{align}
D_{12}&=E\sin\left(\phi - \theta\right),\nonumber\\
D_{23}&=E\sin(\phi + \theta),\\
D_{13}&=\alpha E \sin\phi,\nonumber
\end{align}
where $\theta$ is a bond angle at the base of the triangle.
Here, an additional parameter $\alpha$ is introduced. This parameter reflects the possible difference in local quantum-chemical structural setting responsible for the KNB mechanism \cite{sc_KNB, JJJ, dis}.
The eigenvalues and eigenvectors of the Hamiltonian~\eqref{Ham} are reported in Appendix~\ref{app:A}.

Here, the following notation is adopted for the three-spin basis states, $|\xi_i,\xi_j, \xi_k\rangle=|S_1^z, S_2^z, S_3^z\rangle$.
Interestingly, only two out of eight eigenstates can be ground states at zero temperature for the positive values of magnetic field $B$ and all in-plane directions of the electric field $E$ for the case of antiferromagnetic couplings, $J_1>0$, $J_2>0$. Thus, zero-temperature magnetic behavior of the model is trivial, it has the properties of a two-level system, low-field highly entangled $| \psi_3 \rangle $ with the transition to saturated one $| \psi_1 \rangle $ at
\begin{eqnarray}\label{eq:Bsat}
B_{\text{sat}}=\frac 16 \left(2J_1+J_2-3\sqrt{Q}\cos\frac{\omega+2\pi}{3}\right),
\end{eqnarray}
where the parameters $Q$ and $\omega$ are defined in Appendix through Eq.~\eqref{A2} and the explicit form of all eigenvectors is also presented in Appendix~\ref{app:A}.

\begin{figure}[t]
    \centering
    \includegraphics[width=\linewidth]{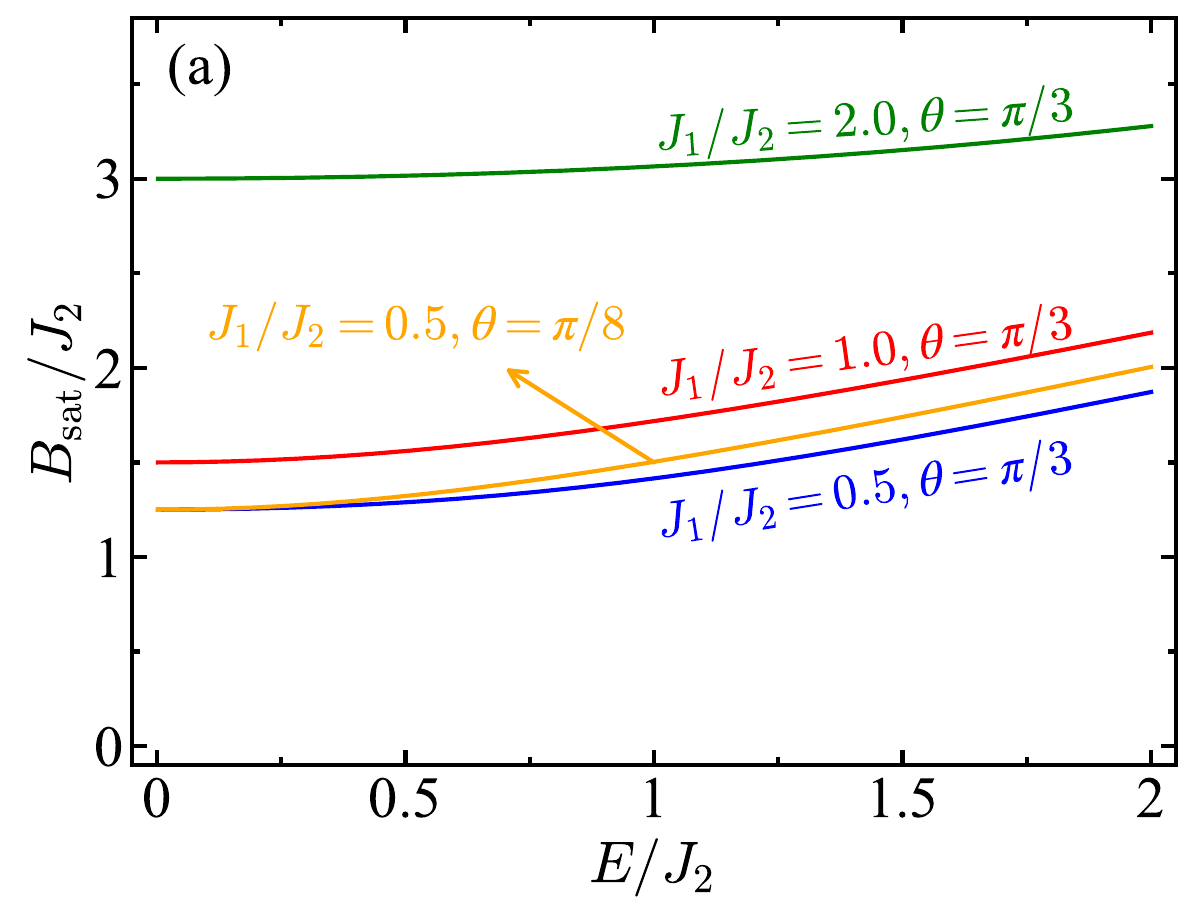}
    \includegraphics[width=\linewidth]{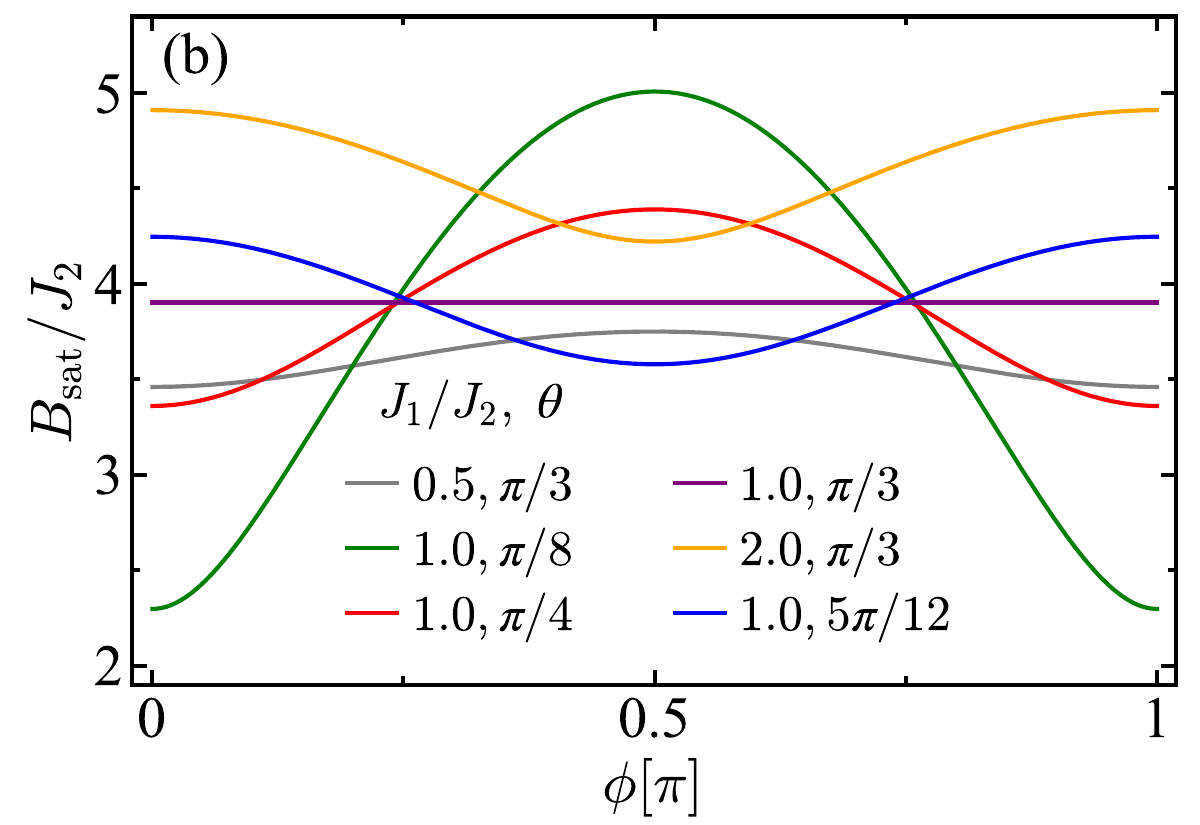}
    \caption{
    Dependence of the saturation magnetic field $B_{\mathrm{sat}}$ on the parameters of the in-plane electric field for the case of uniform KNB coupling ($\alpha=1$). (a) Saturation field as a function of the electric-field magnitude $E/J_2$ for a fixed field orientation, $\phi=\pi/4$.
    The curves are shown for several values of the exchange-coupling ratio $J_1/J_2$ and bond angle $\theta$, illustrating how the electric field modifies the critical magnetic field required to fully polarize the spin trimer. (b) Angular dependence of the saturation field for a rotating electric field of fixed magnitude, $E/J_2=5$, for the same sets of model parameters. The anisotropic variation of $B_{\mathrm{sat}}$ with the electric-field direction reflects the directional character of the KNB spin–electric coupling. The results demonstrate that both the strength and orientation of the external electric field provide efficient means of tuning the saturation field. Moreover, the sensitivity of $B_{\mathrm{sat}}$ to the exchange-coupling ratio $J_1/J_2$ and the bond angle $\theta$ indicates that the magnetic response can be further engineered through the microscopic geometry and magnetic interactions of the molecular nanomagnet.}\label{fig:2}
\end{figure}
The dependence of the saturation field on the magnitude and direction of the electric field is illustrated in Fig. \ref{fig:2} for various values of the angle $\theta$ and the ratio of the exchange constants. One can see a monotonous but relatively slow increase of $B_{\text{sat}}$ with respect to $E$, and more pronounced interplay between the direction of the electric field and the angle of bond $\theta$, which leads to the appearance of the extremum in the graph of the dependence of $B_{\text{sat}}$ on the angle of the electric field $\phi$. The most symmetric case of $\theta=\frac{\pi}{3}$, $J_1/J_2=1$ and $\alpha=1$ is worth separate mentioning, as the value of the saturation field is independent on the electric field angle in this case. This case is remarkable by itself, as the most symmetric one with respect to the rotation of the molecule. Indeed, the Hamiltonian possesses an additional symmetry in this case, namely
\begin{eqnarray}\label{eq:sym}
\mathcal{H}=\mathcal{C}^{-1}\left(2\pi/3\right)\mathcal{H}\mathcal{C}\left(2\pi/3\right),
\end{eqnarray}
where
\begin{eqnarray}\label{eq:sym2}
&&\mathcal{C}\left(\varphi\right)=\mathcal{C}\circ \mathcal{C}_E\left(\varphi\right), \nonumber\\
&& \mathcal{C}\; : \mathbf{S}_i\longrightarrow\mathbf{S}_{i+1},  \\
&& \mathcal{C}_E\left(\varphi\right)\; : \phi\longrightarrow\phi+\varphi.\nonumber
\end{eqnarray}
Thus, the only case when the model with uniform exchange couplings possesses a rotational symmetry is the case of bond angle $\theta=\frac{\pi}{3}$. Only this value can provide an appropriate transformation of the electric field angle, which can compensate the change in DM term under the cyclic permutation of the spin operators in the Hamiltonian. The action of the operator $\mathcal{C}_3$ by itself leads to the following change in the DM coefficient:
\begin{eqnarray}\label{eq:sym3}
D_{12}\;\longrightarrow -D_{13},\quad
D_{13}\;\longrightarrow -D_{23}, \quad
D_{23}\;\longrightarrow D_{12}. \nonumber
\end{eqnarray}
This transformation can be realized by the rotation of the electric field by the angle $\varphi=\frac{2\pi}{3}$ only when $\theta=\frac{\pi}{3}$.

Interestingly, the ground-state properties in zero magnetic field are a bit different; the system exhibits a two-fold degenerate configuration with a coherent mixture of $| \psi_3 \rangle $ and $| \psi_5 \rangle $ eigenstates.
It is also worth mentioning that, there is no continuous limit for all eigenstates and eigenvalues for $E=0$ \cite{ada24,ada24a,ada24b}. The corresponding eigenvalues and the eigenvectors obtained for $E=0$ are provided in Appendix~\ref{app:A}.

The structure of the ground states' dependence on the magnetic field is affected by the ratio of the coupling constants. Here, we consider both couplings as antiferromagnetic. When $J_{1}<J_
{2}$, the system has two ground states  $| \psi_2 \rangle_{0} $ and $| \psi_1\rangle $ with transition at $B_{\text{sat}}=\frac 12 J_1+J_2$.
An additional twofold degeneracy appears at $B=0$ between the ground states $| \psi_2 \rangle_{0} $ and $| \psi_5 \rangle_{0} $.
For $J_{1}>J_{2}$, system again has only two ground states, $| \psi_3 \rangle_{0} $ and  $| \psi_1\rangle $ with additional degeneracy at $B=0$ with  $| \psi_3 \rangle_{0},~| \psi_6 \rangle_{0}$ being the two-fold degenerate zero-field ground states. The saturation field in this case does not depend on $J_2$, i.e. $B_{\text{sat}}=\frac 32 J_1$.
Additionally, some of the eigenvectors do not change continuously under $J_2\rightarrow J_1$. For the case of $J_2=J_1$, when additional symmetry appears, the eigenvectors acquire a new property, chirality, $C^z=\pm 1$, which is the eigenvalue of the scalar chirality operator, or $z$-component of the chirality vector,
\begin{eqnarray}\label{Cz}
C^z=\frac{4}{\sqrt 3}\mathbf{S}_1\left[\mathbf{S}_2\times\mathbf{S}_3\right].
\end{eqnarray}
Thus, as the Hamiltonian in this case commutes with both $C^z$ and $S^z=S_1^z+S_2^z+S_3^z$, the eigenvectors $|\psi_{\left(2,3\right)}\rangle_0$ and $|\psi_{\left(5,6\right)}\rangle_0$ corresponding to $S^z=\pm 1/2$ will take the following forms:
\begin{equation}\label{eq:CS1}
| 1/2, \pm 1 \rangle = \frac{1}{\sqrt{3}} \left(|\downarrow \uparrow \uparrow  \rangle +  q_{\pm}| \uparrow \downarrow\uparrow\rangle+q_{\mp}|\uparrow \uparrow \downarrow \rangle\right),
\end{equation}
and
\begin{equation}\label{eq:CS2}
| -1/2, \pm 1\rangle =\frac{1}{\sqrt{3}} \left(|\uparrow \downarrow \downarrow  \rangle +  q_{\pm}| \downarrow \uparrow\downarrow\rangle+q_{\mp}|\downarrow \downarrow \uparrow \rangle\right),
\end{equation}
respectively, and $q_{\pm}=\exp \left( \pm i2\pi /3\right) $.
For $B=0$, all four of these eigenstates are degenerate.

\section{Quantifying entanglement}
\label{sec3}
To quantify quantum entanglement, various measures can be used~\cite{vid, ami, hor}. For the purposes of the present work, the quantity called negativity is the most convenient~\cite{vid}. As the system under consideration consists of three subsystems (particles), one can speak about bipartite and tripartite entanglement of the eigenstates, which represent a quantitative characteristic of non-separability for the given pair of particles or all three particles, respectively. Let us start with bipartite negativity. Its numerical value varies from 0 (no entanglement) to $1/2$ (maximally entangled pair). For the pair consisting of the particle $i$-th and $j$-th, the negativity, $N_{ij}$, equals the sum of absolute values of negative eigenvalues, $\mu_a$, of the partially transposed reduced two-particle density matrix, $\rho_{ij}^T$, which is constructed in the following way:
\begin{eqnarray}\label{neg1}
&&\left\langle \tilde{\xi_i}, \xi_j \right| \rho_{ij}^T \left| \xi_i, \tilde{\xi_j} \right\rangle= \left\langle \xi_i, \xi_j  \left| \rho_{ij} \right| \tilde{ \xi_i},
\tilde{\xi_j}\right\rangle, \nonumber\\
&&\rho_{ij}=\sum_{\xi_k}\left\langle \xi_k \left| \rho \right|   \xi_k \right\rangle,\;\; k\neq i,j
\end{eqnarray}
where $\left|\xi_i, \xi_j, \xi_k \right\rangle$ is a standard basis defined above. Then, the negativity is obtained according to
\begin{equation}\label{neg2}
N_{ij}=\sum\limits_{a}|\mu_{a}|,
\end{equation}	
where $\mu_a$ stand for the negative eigenvalues of the matrix $\rho_{ij}^{T}$.
Bipartite negativity accounts for the entanglement between pairs of subsystems (particles) within the given eigenstate of the whole system. In order to clarify the overall entanglement of three subsystems, one can use the so-called tripartite negativity. In general, in the system of three particles, three different distributions of the entanglement can be found among its configurations: a fully separable state exhibiting no entanglement; three states where a pair of particles is entangled but the third particle is not (biseparable states); and one configuration in which all three particles are in the entangled state (tripartite entanglement). Tripartite negativity serves as an appropriate measure for characterizing the numerical degree of tripartite entanglement. This measure is defined by the following expression:
\begin{equation}\label{ABC}
N_{ABC}=(N_{A-BC}N_{B-AC}N_{C-AB})^{1/3}
\end{equation}
where $N_{A-BC}, N_{B-AC}, N_{C-AB}$ are generalized bipartite negativities between corresponding particle and the rest of the system~\cite{trip}. This quantity is calculated using formulas similar to those given in Eqs.~\eqref{neg1} and \eqref{neg2}, when the trace in Eq.~\eqref{neg1} is not taken,
{\small \begin{equation}
\left\langle \tilde{\xi_i}, \xi_j, \xi_k \right| \rho_{i-jk}^T \left| \xi_i, \tilde{\xi_j}, \tilde{\xi_k} \right\rangle= \left\langle \xi_i, \xi_j, \xi_k  \left| \rho_{ijk} \right| \tilde{ \xi_i},\tilde{\xi_j}, \tilde{\xi_k}\right\rangle. \nonumber
\end{equation}}
In a spin-(1/2, 1/2, 1/2) Heisenberg trimer, the negativity varies between zero and $1/2$. Since our analysis is restricted to quantum entanglement in the ground-state regime (zero-temperature entanglement), the system is described by pure states only. Therefore, the density matrix $\rho$ corresponding to each of the eight eigenstates of the Hamiltonian is defined as a pure-state projector, given by
\begin{equation}
\rho_{i}=|\Psi_{i}\rangle \langle \Psi_{i}|, \;\; i=1,...,8.
\end{equation}
In the case of $n$ degenerate eigenstates, one should use
\begin{eqnarray}\label{eq: rho_degenerate}
\rho_{i_1...i_n}=\frac {1}{n}\sum_{a=1}^n|\Psi_{i_a}\rangle \langle \Psi_{i_a}| .
\end{eqnarray}

For the system under consideration, all bipartite and tripartite negativities can be calculated analytically. The highly symmetric situations, for which the continuous-limit treatment is not applicable, must be analyzed separately. This includes the cases of degenerate ground states, such as at the saturation magnetic field or when both the magnetic and electric fields vanish ($B=0$ and $E=0$), where the density matrix given by Eq.~(\ref{eq: rho_degenerate}) should be employed. We first consider the most symmetric configuration characterized by homogeneous exchange couplings ($J_1=J_2>0$) in the absence of external magnetic and electric fields. In this regime, the ground state exhibits a fourfold degeneracy and is spanned by the eigenstates defined in Eqs.~\eqref{eq:CS1} and \eqref{eq:CS2}. Remarkably, although all pairwise negativities vanish for these states, the tripartite negativity remains finite,
\begin{eqnarray}
N_{12}=N_{13}=N_{23}=0, \; N_{123}=\frac 16,
\end{eqnarray}
revealing the presence of genuine multipartite quantum entanglement that cannot be captured by bipartite entanglement measures alone.

For the case of non-zero magnetic field below saturation, the degeneracy is lifted and now only two eigenvectors with $S^z=1/2$ correspond to the ground state. So, the bipartite and tripartite negativities are given by
\begin{eqnarray}
&&N_{12}=N_{13}=N_{23}=\frac{1}{6} (\sqrt{2} - 1)\approx0.07, \nonumber\\
&&N_{123}=\frac{1}{3 \sqrt{2}}\approx0.24.
\end{eqnarray}
Next, we consider the more general case of zero-field ground states, i.e.   $0<J_{1}<J_{2}$ and $0<J_{2}<J_{1}$. For the dominant antiferromagnetic $J_2$ and $B=0$, the ground state is presented by the degenerate doublet states $| \psi_2 \rangle_{0} $ and $| \psi_5 \rangle_{0} $. Both of these states are biseparable, which means that the corresponding pure-state density matrix is factorizable $\rho_{12}\otimes\rho_{3}$, where the indices denote the number of the sites in the triangle. Thus, bipartite negativities are
\begin{eqnarray}
N_{12}^{(2+5)_{0}}=\frac{1}{2}, \; N_{13}=N_{23}=0
\end{eqnarray}
For the non-zero magnetic field below saturation, the degeneracy is lifted but the entanglement properties remain mostly the same,
\begin{eqnarray}
N_{12}^{2_{0}}=\frac{1}{2}.
\end{eqnarray}
For $0<J_{2}<J_{1}$ case, again the zero-field ground state is a degenerate doublet,  $| \psi_3 \rangle_{0} $ and $| \psi_6 \rangle_{0} $, which transforms to  $| \psi_3 \rangle_{0} $ after the lifting of the degeneracy by the finite magnetic field below saturation. The corresponding bipartite and tripartite negativities then read:
 \begin{eqnarray} &&N_{12}^{(3+6)_{0}}=N_{23}^{(3+6)_{0}}= \frac{1}{4}, \nonumber\\
&&N_{13}^{(3+6)_{0}}= 0,  \nonumber\\
&&N_{1-23}^{(3+6)_{0}}=N_{3-12}^{(3+6)_{0}}= \frac{1}{6} \left(\sqrt{13}-2\right)\approx0.27, \nonumber\\
&&N_{2-13}^{(3+6)_{0}}= \frac{1}{3},  \nonumber\\
&&N_{123}^{(3+6)_{0}}=\frac 13 \left(\frac 12\left(\sqrt{13}-2\right)\right)^{2/3}\approx0.29.\nonumber
\end{eqnarray}
and for $B\ne0$, we have
\begin{eqnarray}
      &&N_{12}^{3_{0}}=N_{23}^{3_{0}}= \frac{1}{12} \left(\sqrt{17}-1\right)\approx0.26,  \nonumber\\
      &&N_{13}^{3_{0}}= \frac{1}{6} \left(\sqrt{5}-2\right) \approx0.04, \nonumber \\
      &&N_{1-23}^{3_{0}}=N_{3-12}^{3_{0}}= \frac{\sqrt{5}}{6}  \approx0.37,   \nonumber\\
      &&N_{2-13}^{3_{0}}= \frac{\sqrt{2}}{3}  \approx0.47, \nonumber\\
      &&N_{123}^{3_{0}}=\frac 13\left(\frac{5}{2}\right)^{1/6} \approx 0.39. \nonumber
\end{eqnarray}
In the general case ($B\ne0$ and $E\ne0$), expressions for bipartite negativities and components of $N_{123}^3$ for the non-trivial ground state, $|\psi_3\rangle$, are given by
\begin{eqnarray}\label{eigval}
&&N_{12}^3=\frac{\sqrt{\mid L_{1}\mid^4+4\mid M_{1}\mid^2}-\mid L_{1}\mid^2}{2(1+\left| L_{1}\right| ^2+\left| M_{1}\right| ^2)},\nonumber\\
    &&N_{13}^3=\frac{\sqrt{\mid L_{1}\mid^2+4\mid M_{1}\mid^4}-\mid L_{1}\mid^2}{2(1+\left|L_{1}\right| ^2+\left| M_{1}\right| ^2)}, \nonumber\\
    &&N_{23}^3=\frac{\sqrt{1+4\mid L_{1}\mid^2\mid M_{1}\mid^2}-1}{2(1+\left|L_{1}\right| ^2+\left| M_{1}\right| ^2)}, \nonumber\\
    &&N_{1-23}^3=\frac{\sqrt{\left| L_{1}\right|^2+\left| M_{1}\right|^2}}{1+\left| L_{1}\right|^2+\left| M_{1}\right| ^2}, \nonumber\\
    &&N_{2-13}^3=\frac{\left| L_{1}\right| \sqrt{1+\left| L_{1}\right| ^2} }{1+\left|
    L_{1}\right| ^2+\left| M_{1}\right| ^2}, \nonumber \\
    &&N_{3-12}^3=\frac{\left| M_{1}\right| \sqrt{1+\left| M_{1}\right| ^2} }{1+\left|
    L_{1}\right| ^2+\left| M_{1}\right| ^2},
\end{eqnarray}
where $L_1$ and $M_1$ are given in Appendix~\ref{app:A}.
 The role of the magnetic field here is limited to the external switch, which can be used to drive the system into a nonentangled saturated state or to keep in a non-saturated entangled state. As the magnetization is conserved here, in contrast to the systems with non-conserved magnetization~\cite{ada20, ada24, ada24a, ada24b}, entanglement within the same eigenstate has no dependence on the magnetic field. However, as the terms describing the interaction between the KNB polarization and in-plane electric field (Eq.~\eqref{eq: KNB_pol}) do not commute with the Hamiltonian, entanglement properties of the model can be efficiently manipulated by the electric field.

\section{Bipartite entanglement transfer in rotating electric field}
\label{sec4}

\begin{figure}[t]
    \centering
     \includegraphics[width=1\linewidth]{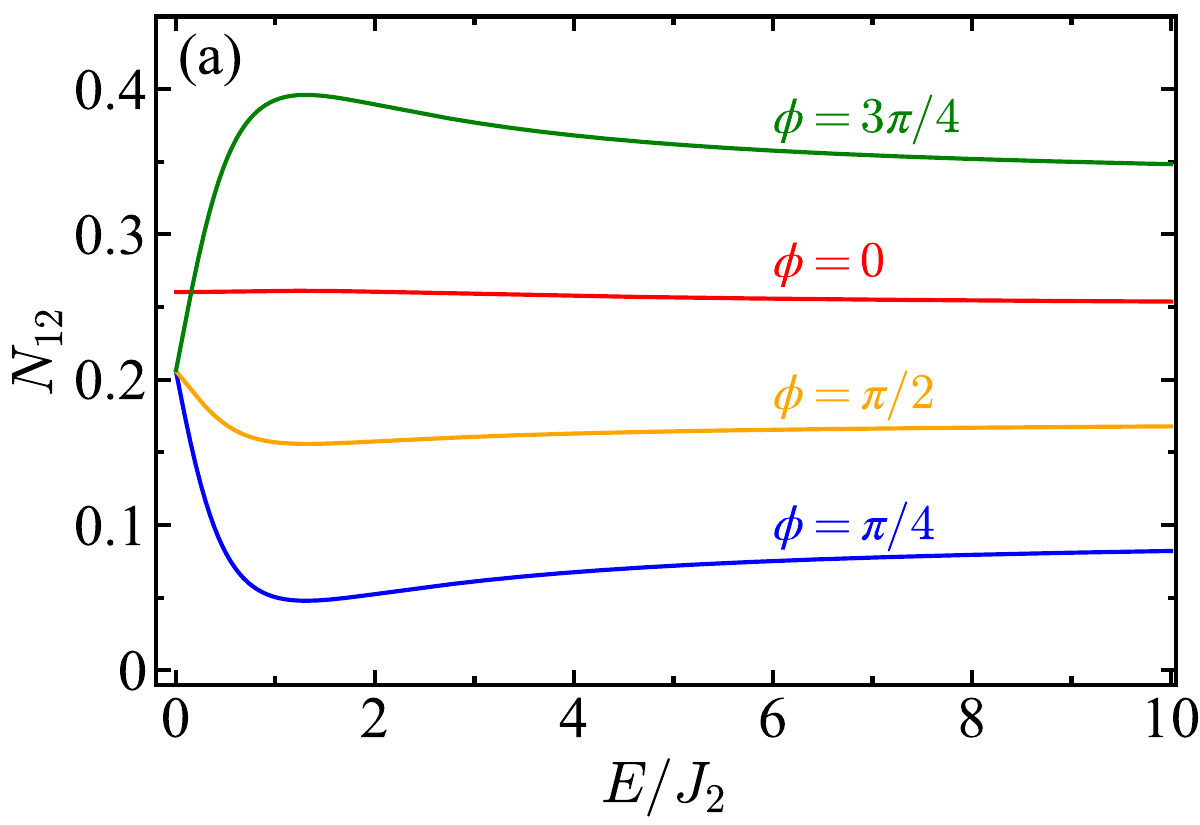}
    \includegraphics[width=1\linewidth]{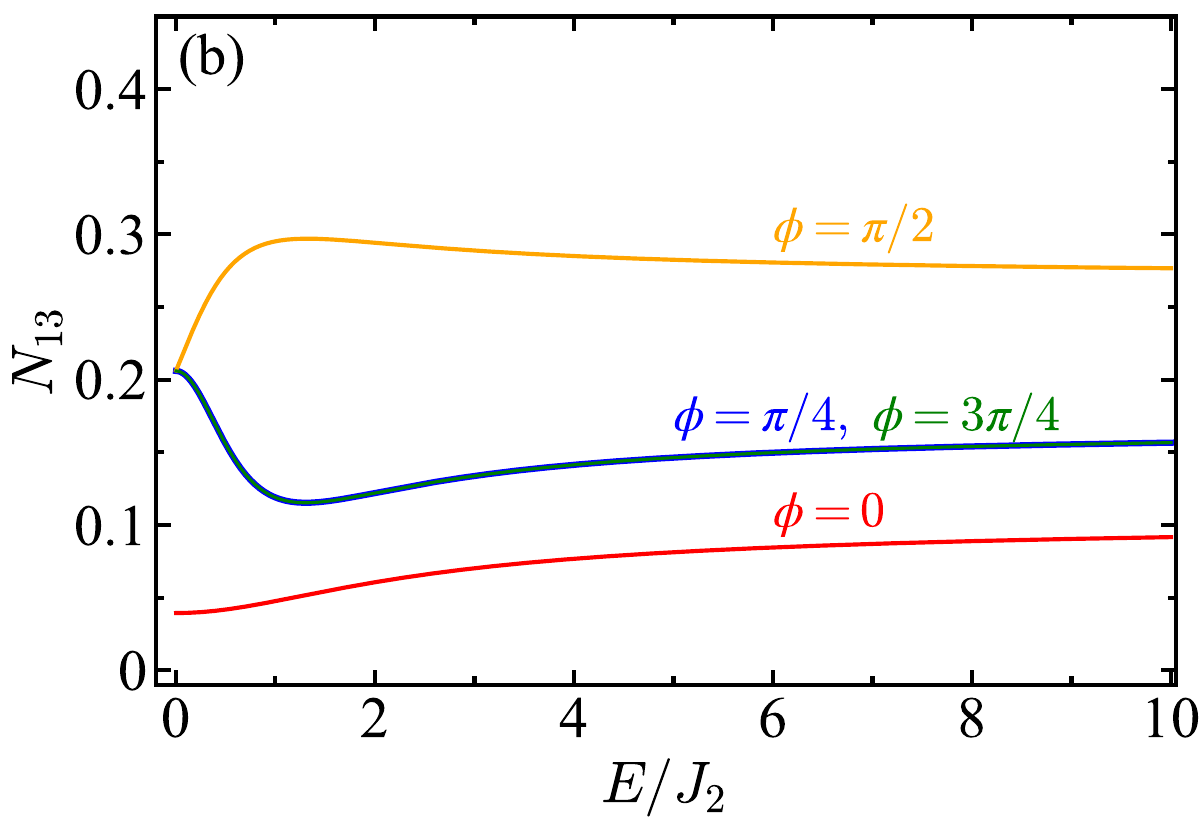}
    \includegraphics[width=1\linewidth]{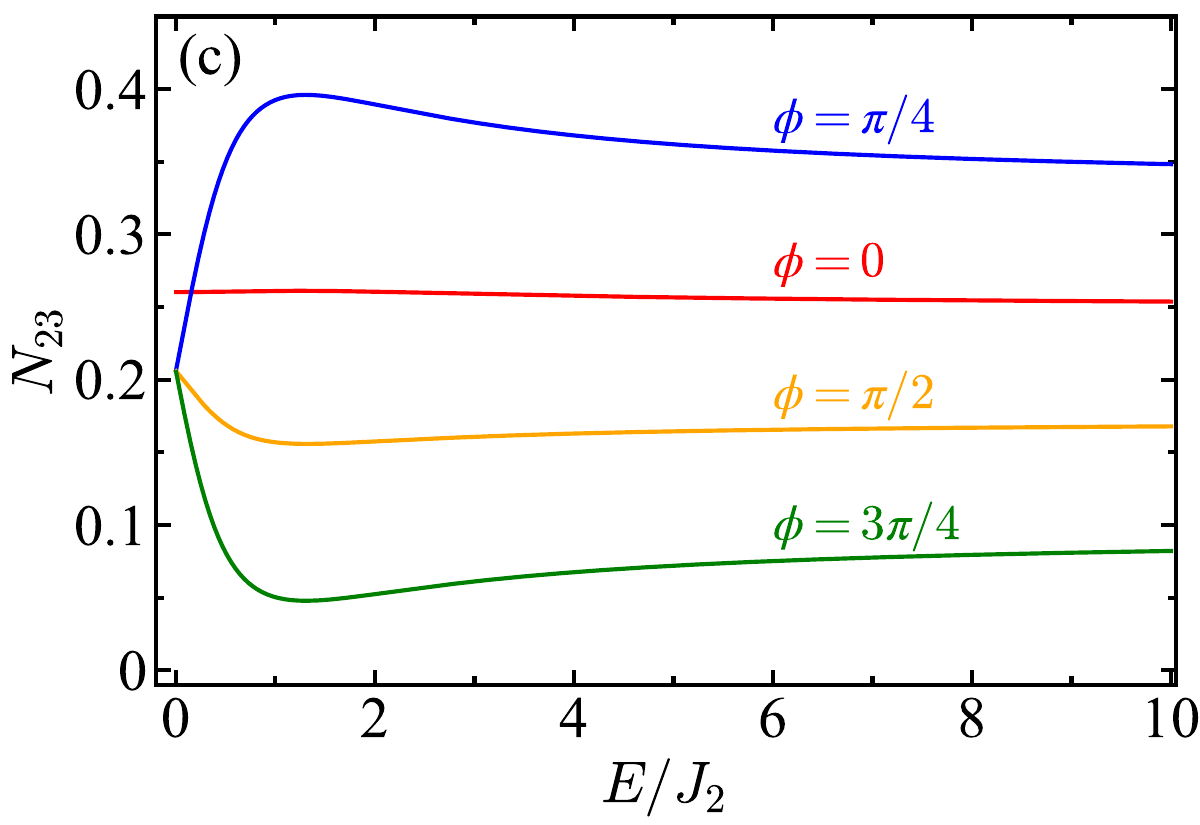}
    \caption{Bipartite negativity dependence on the magnitude of electric field, $E$, with $J_1=J_2=1$, $\alpha=1$ and $\theta=\pi/4$  for four different values of the electric field angle, $\phi=0,~\pi/4,~\pi/2$ and $3\pi/4$. 
    }
    \label{fig:3}
\end{figure}

As discussed in the previous section, the KNB mechanism provides a unique opportunity to manipulate the entanglement properties of magnetic materials. An external electric field can be used either to enhance or to suppress the quantum entanglement between atomic spins within a magnetic molecule. Owing to the bond-dependent action of the electric field inside the same molecule, a consequence of both the molecular geometry and the orientation of the electric field (see the KNB polarization expression in Eq. (\ref{eq: KNB_pol})), the resulting effect may differ significantly for different spin pairs. Therefore, in the proposed geometry, where the electric field lies in the plane of the spin triangle, one can expect a nonuniform electric-field dependence of the bipartite negativities corresponding to different spin pairs. Moreover, the pronounced sensitivity not only to the magnitude of the electric field but also to its direction makes this mechanism a highly efficient tool for controlling the local entanglement properties of the magnetic molecule.
As an illustration of the non-homogeneous action of the electric field on different spin pairs, the plots of bipartite entanglement dependence for three spin pairs and for several values of the electric field angle $\phi$ are presented in Fig. \ref{fig:3}. These plots demonstrate the symmetry between $N_{12}$ and $N_{23}$ which is the consequence of the problem geometry, $N_{12}\left(E, \phi\right)=N_{23}\left(E, \phi\pm\pi\right)$ for $\phi \neq 0, \pm \pi, \pm\pi/2$ and $N_{12}\left(E, 0\right)=N_{23}\left(E, 0\right)$, $N_{12}\left(E, \pi/2\right)=N_{23}\left(E, \pi/2\right)$. For electric-field orientations corresponding to $\theta=\pi/4$ and $\theta=3\pi/4$, the electric field exerts opposite effects on the entanglement of the spin pairs $(1,2)$ and $(2,3)$. Specifically, an increase in the electric-field strength enhances the negativity $N_{12}$ while simultaneously decreasing $N_{23}$, and vice versa. This behavior highlights the redistribution of bipartite entanglement induced by the electric field between neighboring spin pairs. For $\phi=\pi/2$, the negativity $N_{13}$ exhibits a trend opposite to those of $N_{12}$ and $N_{23}$. Furthermore, the dependence of $N_{13}$ on the electric-field strength is identical for $\phi=\pi/4$ and $\phi=3\pi/4$. Despite these differences, all curves presented in Fig.~\ref{fig:3} share the same characteristic features: a pronounced variation at low electric-field strengths, followed by the emergence of a local minimum or maximum, and finally a slow asymptotic convergence toward a constant value at larger electric-field magnitudes.

\begin{figure}[t]
    \centering
    \includegraphics[width=1\linewidth]{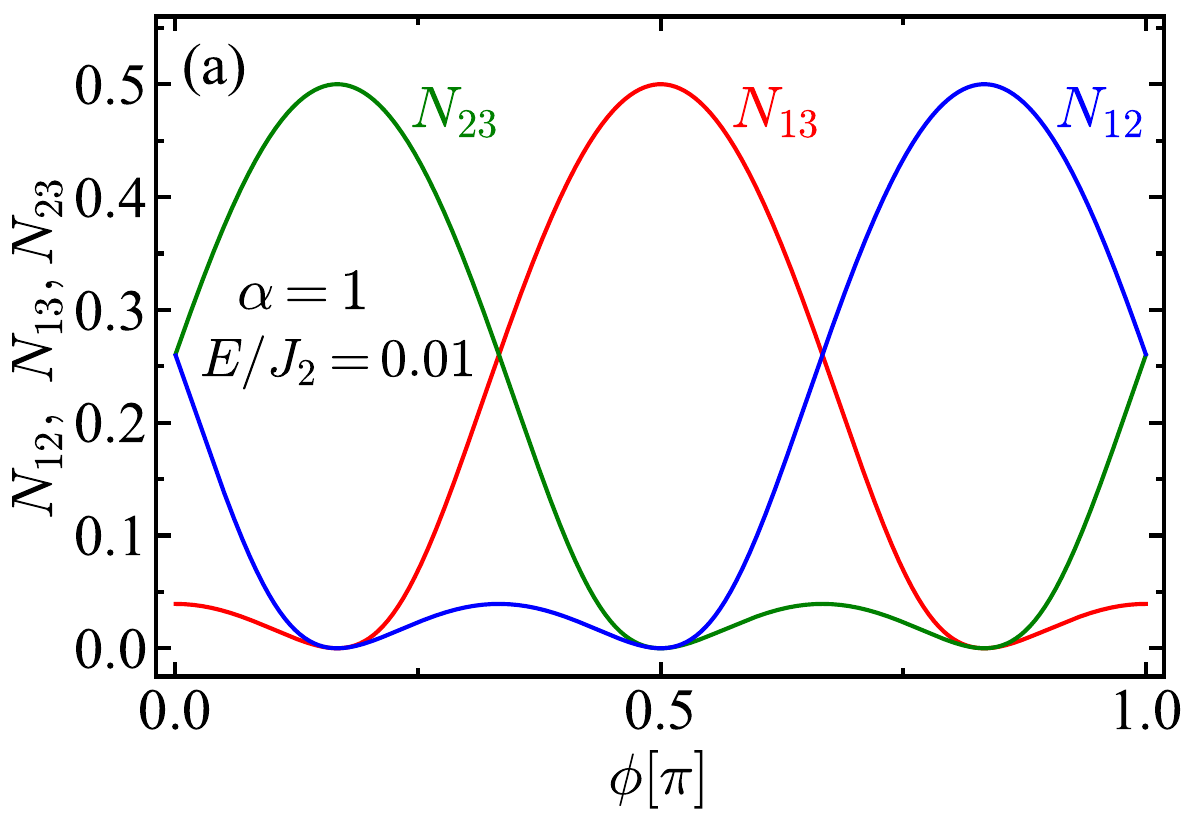}
    \includegraphics[width=1\linewidth]{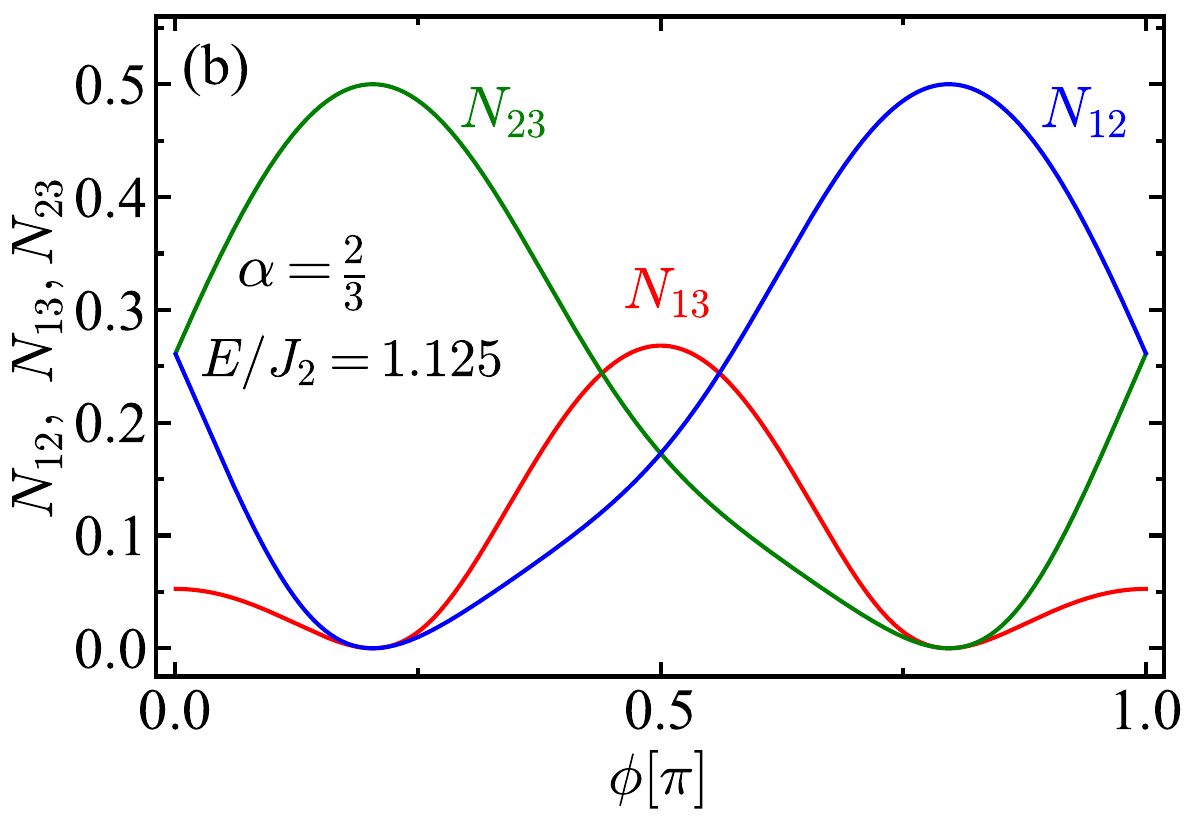}
    \includegraphics[width=1\linewidth]{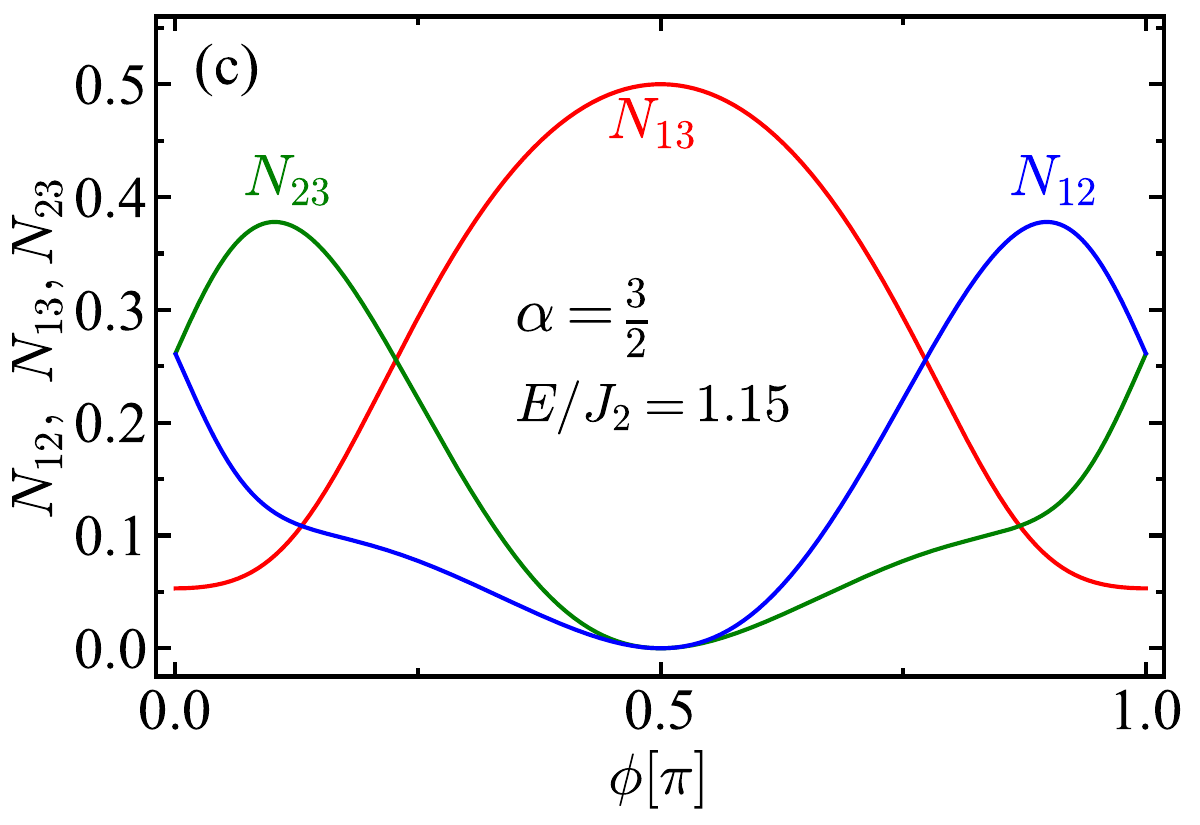}
    \caption{Bipartite negativity dependence on the direction of electric field, $\phi$, with $J_1=J_2=1$ and $\theta=\pi/3$. Three panels present $N_{ij}$ for different values of  $\alpha$ and electric field magnitude. Panel (a): $\alpha=1$,  $E/J_2=0.01$; panel (b): $\alpha=2/3$, $E/J_2=1.125$; and panel (c): $\alpha= 3/2$, $E/J_2=1.15$.  The blue, red and green lines correspond to $N_{12}$, $N_{13}$ and $N_{23}$ respectively. }\label{fig:4}
\end{figure}

The most intriguing behavior of the bipartite entanglement can be observed when the electric field vector, remaining constant in magnitude, is rotating in the plane of the spin triangle. Studying the magneto-electric response of various models in the setup of a rotating sample in a homogeneous electric field is often referred to as RMEE~\cite{cen19, cen21}. Concerning the local entanglement properties, the design of this experiment provides an additional advantage, a possibility of controlled transfer of the bipartite entanglement between the pair of system spins. The dependence of bipartite negativities on the electric field angle $\phi$ is demonstrated in Fig. \ref{fig:4} for the most symmetric case, $J_1=J_2=1$ and $\theta=\pi/3$. Due to obvious symmetry, these dependencies have a period equal to $\pi$.  Three panels correspond to different values of the parameter $\alpha$, which can now serve as a symmetry-breaking factor. The most symmetric case, $\alpha=1$, presented in panel (a), shows an almost perfect transfer of the entanglement between three pairs of spins in a rotating electric field. The notion of perfect transfer refers to the situation in which the bipartite negativity of a given pair reaches its maximal value $N_{ij}=1/2$. In this case, the negativities of the other two pairs become exactly zero, demonstrating a complete redistribution of entanglement such that all bipartite quantum correlations are concentrated within a single pair.
However, the perfect transfer of bipartite negativity is not possible, as it implies the appearance of the perfect biseparable eigenstate of the three spins at certain value of the electric field angle, in which two spins form the spin singlet and the third one is separable:
  \begin{eqnarray}
 \left|\psi\right\rangle=\frac{1}{\sqrt 2}\left(\left|\uparrow_i\downarrow_j\right\rangle-\left|\downarrow_i\uparrow_j\right\rangle\right)\bigotimes\left|\xi_k\right\rangle,
  \end{eqnarray}
where $\left|\xi_k\right\rangle$ can be either $\left|\uparrow\right\rangle$ or $\left|\downarrow\right\rangle$. This can not be achieved for the non-trivial ground state,
\begin{eqnarray}
&&\left|\psi_3\right\rangle\sim
C_1\left|\downarrow\uparrow\uparrow\right\rangle+C_2\left|\right\uparrow\downarrow\uparrow\rangle+C_3\left|\uparrow\uparrow\downarrow\right\rangle,
\end{eqnarray}
with
\begin{eqnarray}
&&C_1=\frac{1}{\sqrt{1+|L_1|^2+|M_1|^2}}, \nonumber\\
&&C_2=\frac{M_1}{\sqrt{1+|L_1|^2+|M_1|^2}}, \nonumber\\
&&C_3=\frac{L_1}{\sqrt{1+|L_1|^2+|M_1|^2}}, \nonumber
\end{eqnarray}
as none of the coefficients $C_1$, $C_2$, and $C_3$ can be made equal to zero for any finite value of $E$ and $\phi$.

Nevertheless, the negativity profiles shown in Fig.~\ref{fig:4}(a) exhibit an almost perfect transfer of entanglement. The maximum and minimum values of the negativities deviate from the values of $1/2$ and $0$, respectively, by only $10^{-8}$ to $10^{-10}$. Such tiny discrepancies are negligible for all practical purposes and indicate an essentially complete transfer of bipartite entanglement between the spin pairs.
The maximal value of the bipartite negativity for each pair of spins is achieved when the electric field vector is normal to the bond connecting these spins. Thus, the maximal entanglement can be transferred to different spin pairs by tuning the electric-field angle. Specifically, $N_{12}=1/2$ at $\phi=5\pi/6$, $N_{13}=1/2$ at $\phi=\pi/2$, and $N_{23}=1/2$ at $\phi=\pi/6$. This behavior illustrates the electric-field-induced steering of bipartite entanglement within the spin trimer.

Panels (b) and (c) of Fig.~\ref{fig:4} display a similar entanglement-transfer behavior; however, the maximum negativity values become unevenly distributed due to the nonhomogeneous microscopic KNB properties of the bond characterized by $\alpha \neq 1$. Fig.~\ref{fig:4}(b) corresponds to the case $\alpha < 1$, where the coupling between the electric field and the spin pair $1-3$ is weaker than that associated with the other two pairs. Although the overall qualitative picture of entanglement transfer remains unchanged, the maximum value attained by $N_{13}$ is noticeably reduced, reflecting the weaker electric-field control over this bond.  Small shifts in the electric-field angles corresponding to the maxima of $N_{12}$ and $N_{23}$ are also observed.

The opposite situation is shown in Fig.~\ref{fig:4}(c), which corresponds to $\alpha>1$. In this case, the coupling between the electric field and the spin pair
 $1-3$ is stronger than that of the other two pairs, leading to a relative suppression of the maximum values of $N_{12}$ and $N_{23}$, accompanied by peak shifts in the opposite direction. In contrast, the position of the $N_{13}$ maximum remains unaffected by the nonuniform magnetoelectric coupling and is always located at $\phi=\pi/2$.
It is also worth noting that, owing to the underlying symmetries of the system, the negativity $N_{13}$ is always symmetric with respect to $\phi=\pi/2$. Furthermore, the negativities $N_{12}$ and $N_{23}$ are related by mirror symmetry about the same angle, $\phi=\pi/2$, such that their angular dependences are reflections of one another.


\begin{figure}[t]
    \centering
    \includegraphics[width=1\linewidth]{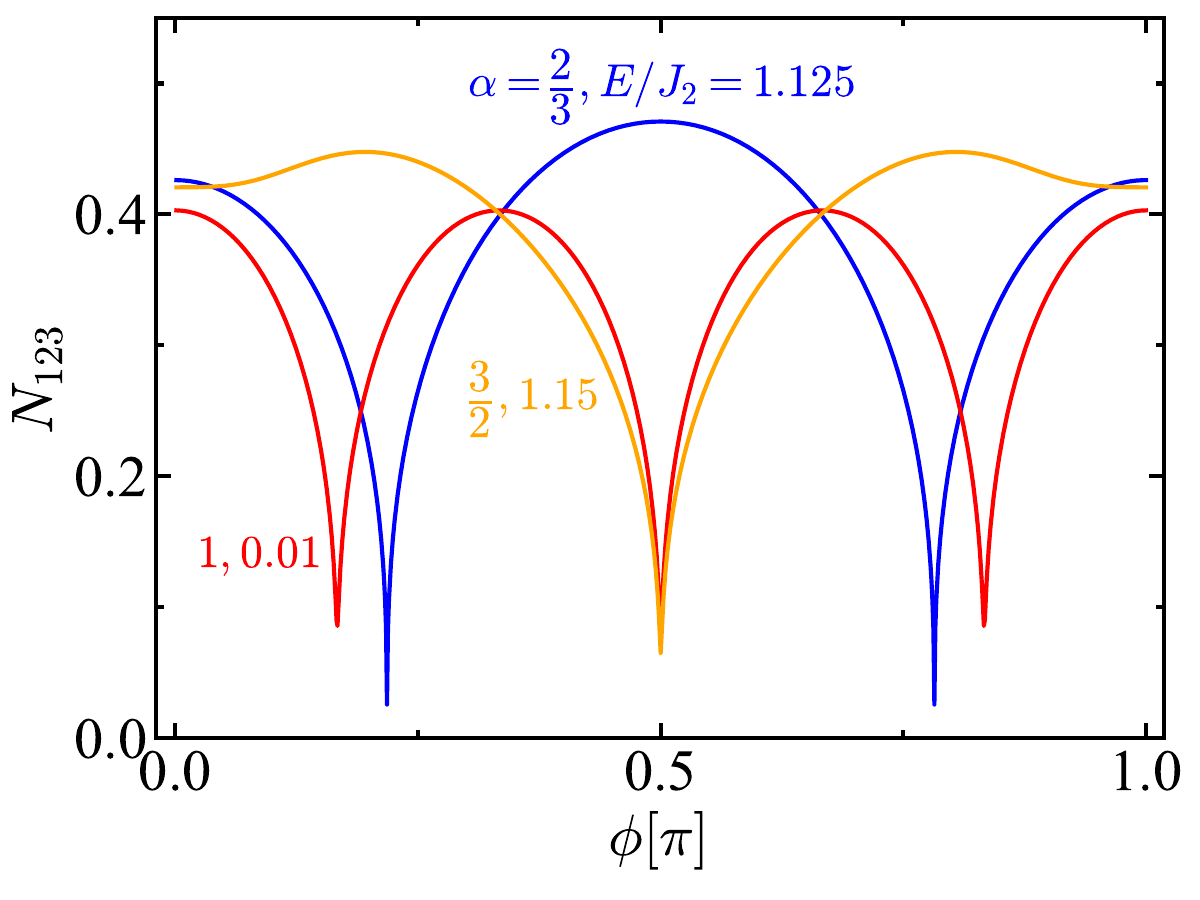}
    \caption{Tripartite negativity dependence on the direction of electric field $\phi$ for the three particular cases, which correspond to the parameter sets used in Fig. \ref{fig:4}(a): red curve; \ref{fig:4}(b): blue curve; and  \ref{fig:4}(c): orange curve.}\label{fig:5}
\end{figure}

The tripartite negativity, which is shown in Fig.~\ref{fig:5} for the same sets of parameters as used in Fig.~\ref{fig:4}, is also periodic under the action of the rotating electric field. The most intriguing features observed in the tripartite negativity are pronounced and sharply localized minima. Depending on the value of $\alpha$, the number of these minima can be one, two, or three, revealing a strong dependence of the multipartite entanglement on the degree of nonuniformity in the magnetoelectric coupling.

\begin{figure}[t]
    \centering
    \includegraphics[width=1\linewidth]{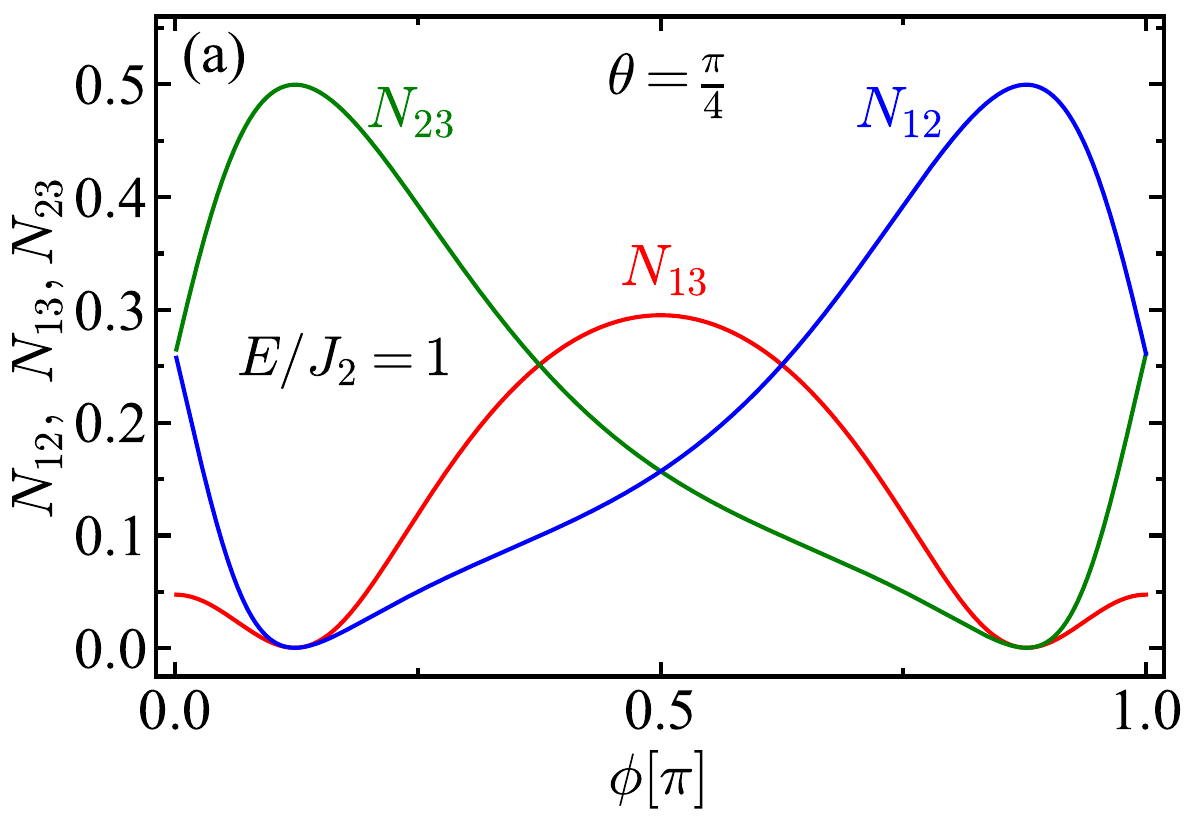}
    \includegraphics[width=1\linewidth]{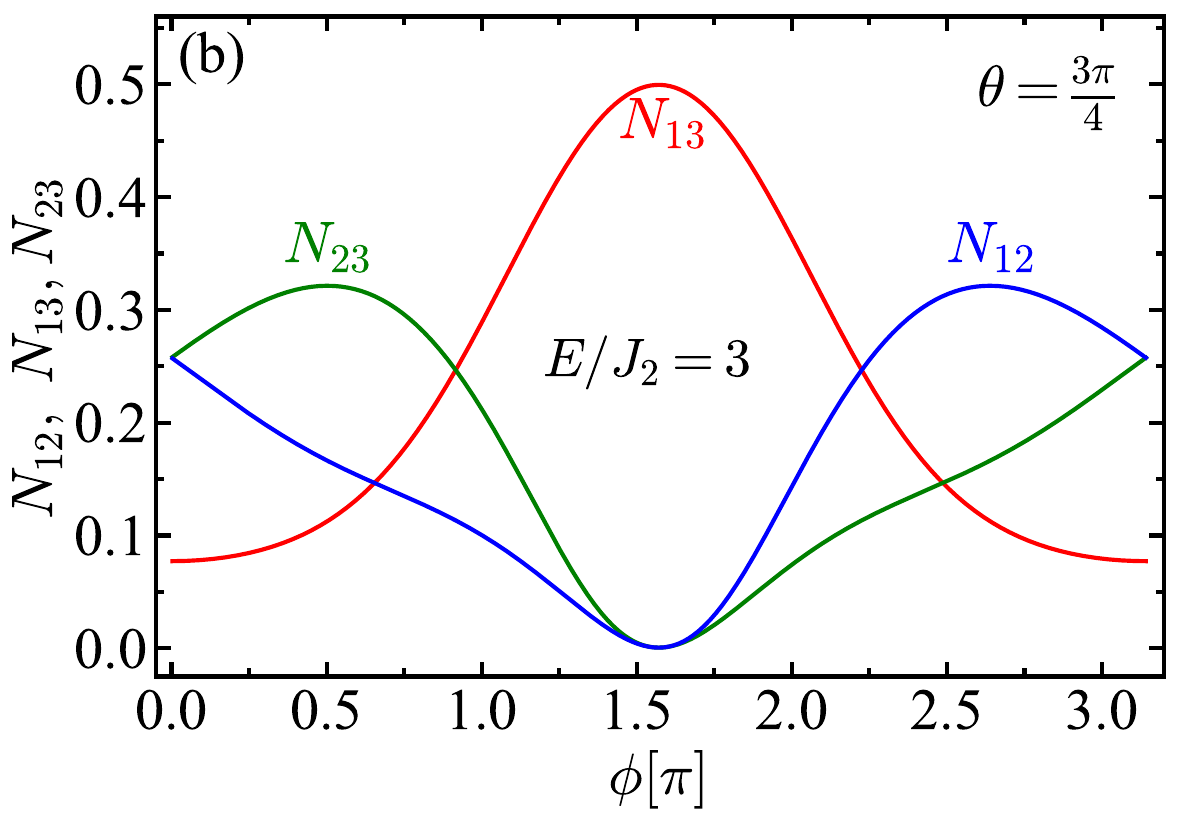}
    \caption{Bipartite negativity dependence on the direction of electric field, $\phi$, for $J_1=J_2=1$ and $\alpha=1$. (a) $\theta=\pi/4$  and (b) $\theta=3\pi/4$.  The blue, red and green lines correspond to $N_{12}$, $N_{13}$ and $N_{23}$, respectively. }\label{fig:6}
\end{figure}

\begin{figure}[t]
    \centering
    \includegraphics[width=1\linewidth]{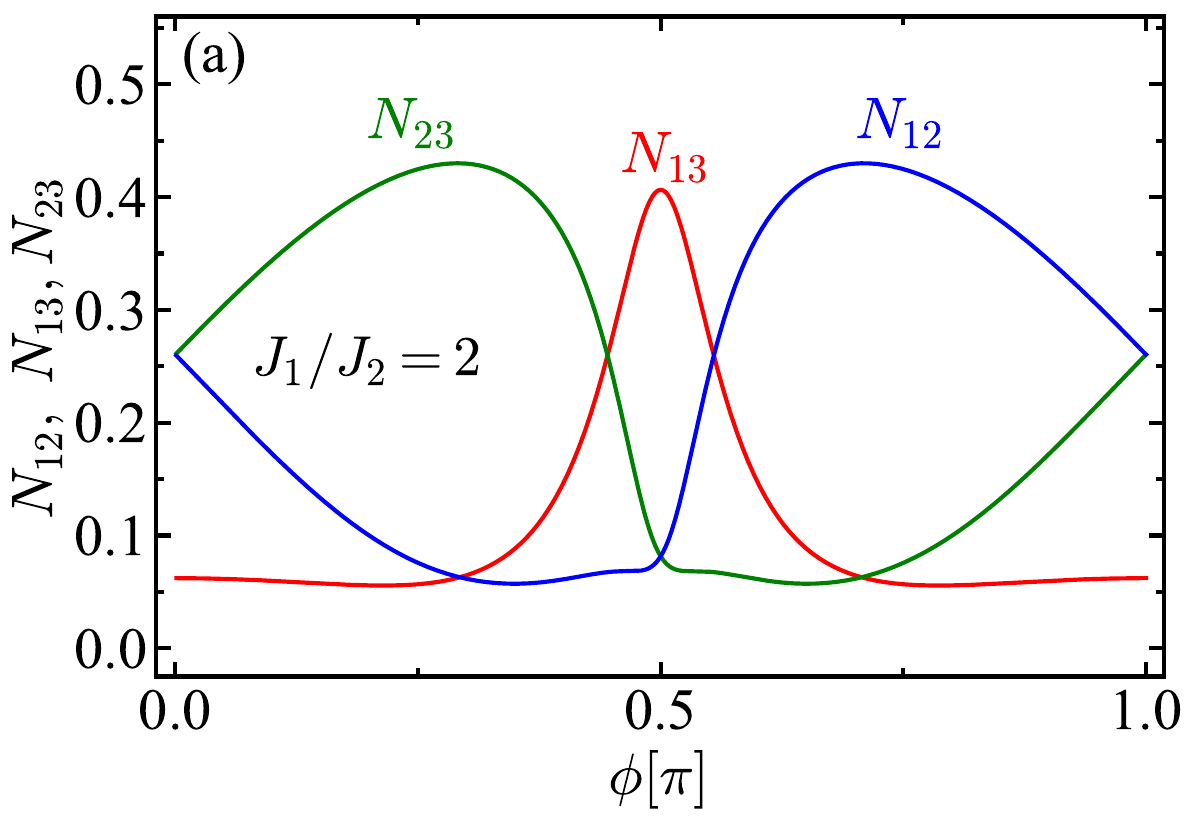}
    \includegraphics[width=1\linewidth]{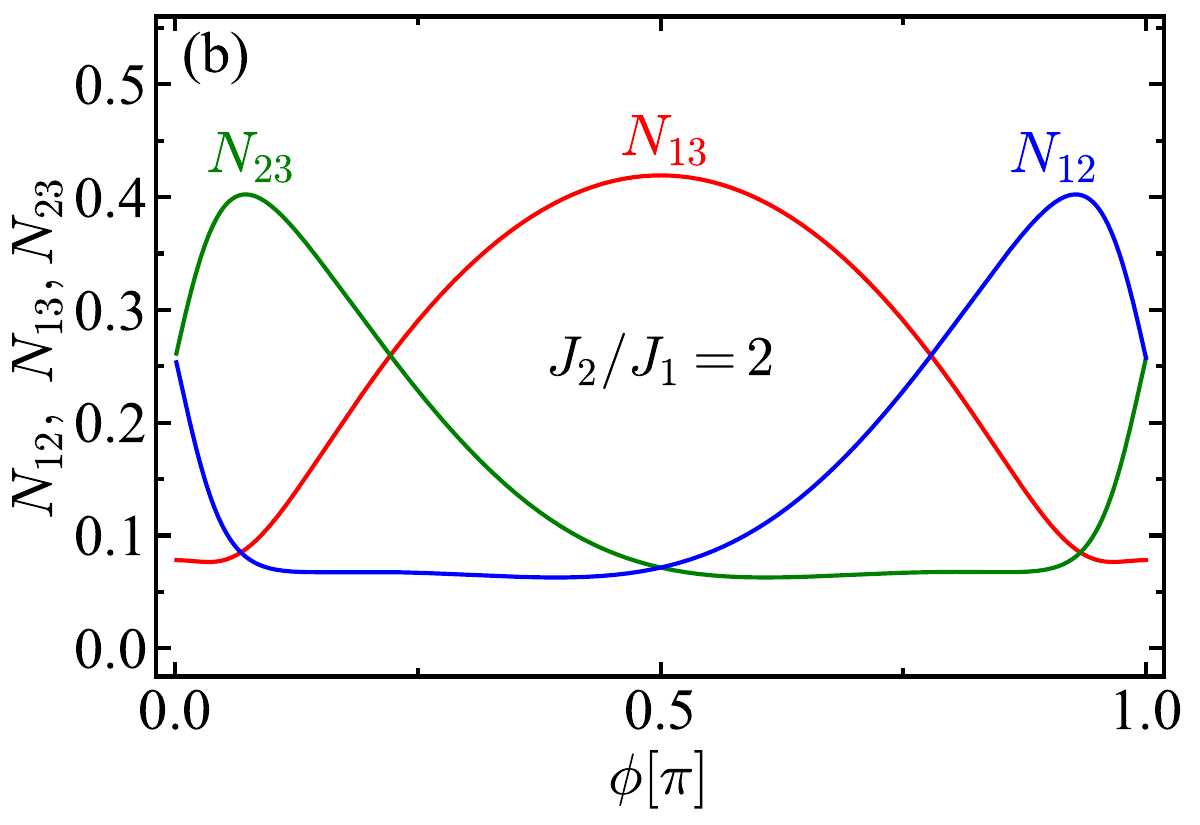}
    \caption{Bipartite negativity dependence on the direction of electric field, $\phi$, for $\theta=\pi/3$ and $\alpha=1$. (a) $J_1/J_2=2$ and $E/J_2=3.5$. (b) $J_2/J_1=2$ and  $E/J_1=3.5$. The blue, red and green lines correspond to $N_{12}$, $N_{13}$ and $N_{23}$, respectively. }\label{fig:7}
\end{figure}
Note that the modification of the entanglement-transfer ratio induced by the parameter $\alpha$, as shown in panels (b) and (c) of Fig.~\ref{fig:4}, can likewise be realized by adjusting the bond angle $\theta$ and/or the interaction ratio $J_2/J_1$. This demonstrates that comparable control over the redistribution of bipartite entanglement may be achieved through either the microscopic magnetoelectric coupling or the geometric and magnetic interaction parameters of the system. For $\theta < \pi/3$ and $\theta > \pi/3$, the entanglement-transfer behavior under a rotating electric field is qualitatively similar to that shown in Fig.~\ref{fig:4}(b) and Fig.~\ref{fig:4}(c), respectively (see Figs.~\ref{fig:6}(a) and \ref{fig:6}(b)). A similar correspondence is observed when a relative strength of the coupling constants is varied. Specifically, the cases $J_2<J_1$ and $J_2>J_1$, illustrated in Figs.~\ref{fig:7}(a) and \ref{fig:7}(b), respectively, exhibit the same qualitative features of entanglement transfer. These results indicate that tuning either the bond angle $\theta$ or the relative strength of the coupling constants $J_2/J_1$ produces effects analogous to those arising from variations in the parameter $\alpha$.

\begin{figure*}[t]
    \includegraphics[width=1\linewidth]{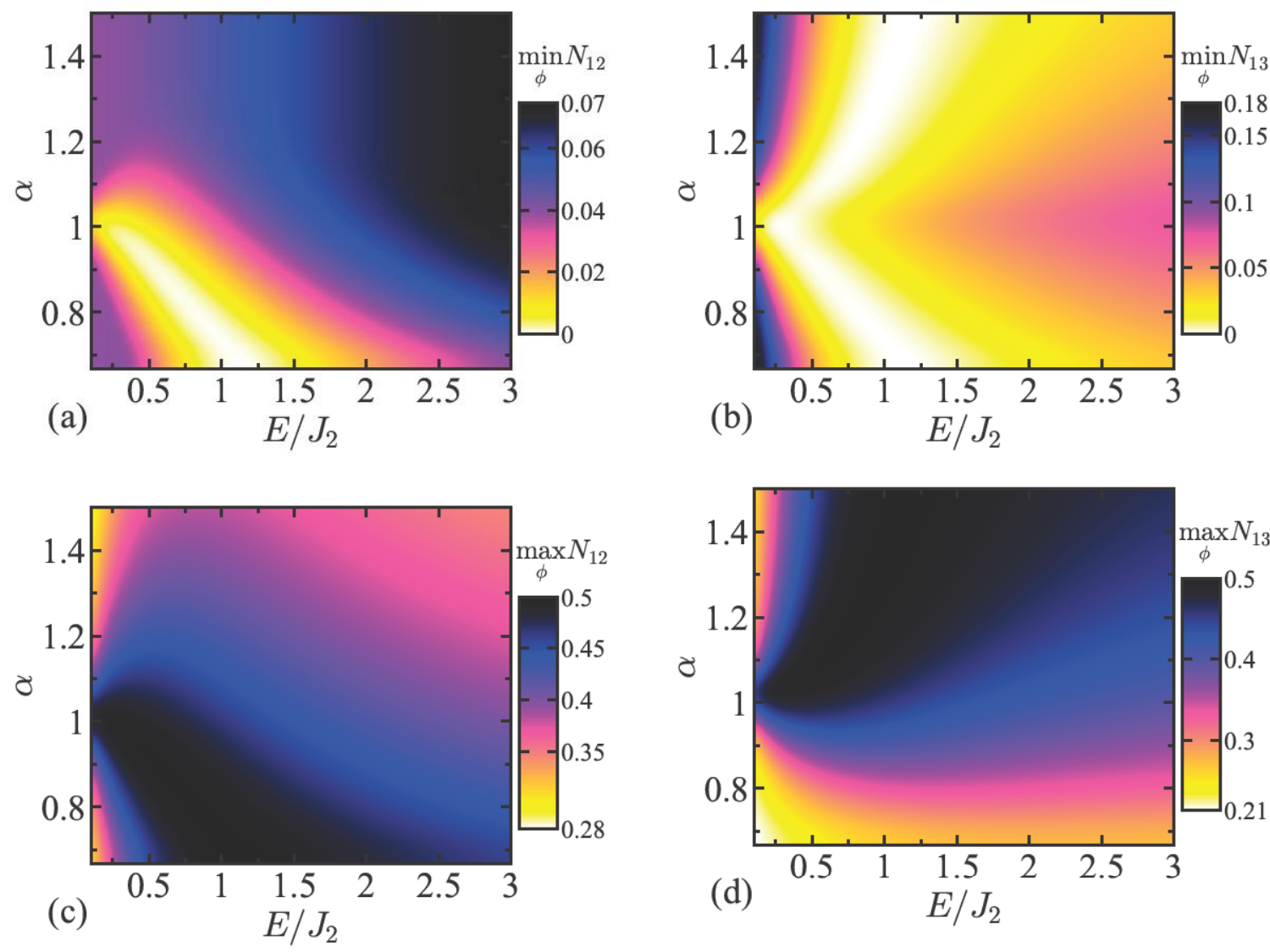}
    \caption{Density plots of the minimum (panels (a) and (b)) and maximum (panels (c) and (d)) values of the bipartite negativity attained during a complete rotation of the in-plane electric field ($0\leq\phi\leq\pi$). Panels (a) and (c) correspond to the negativity $N_{12}$ ($=N_{23}$ by symmetry), whereas panels (b) and (d) show the corresponding extrema of $N_{13}$. The calculations are performed for homogeneous exchange interactions ($J_1=J_2=1$) and an equilateral triangular geometry ($\theta=\pi/3$), with the results presented in the ($\alpha,~E/J_2$) parameter plane. The density plots provide a comprehensive overview of the range of bipartite entanglement accessible through electric-field control, illustrating how the field strength and the nonuniformity of the KNB spin–electric coupling jointly determine the suppression and enhancement of pairwise quantum correlations. Regions where the maximum negativity approaches its theoretical upper bound ($N_{ij}=1/2$) while the corresponding minimum tends to zero identify parameter regimes in which the rotating electric field enables an almost ideal transfer of bipartite entanglement between different spin pairs. In contrast, deviations from these regions reveal the influence of nonuniform spin–electric coupling on the efficiency of the entanglement-transfer process and the redistribution of quantum correlations within the molecular nanomagnet.
    }\label{fig:8}
\end{figure*}

A comprehensive overview of the bipartite entanglement redistribution during the rotation of the electric field from $\phi=0$ to $\phi=\pi$ is presented in Fig.~\ref{fig:8}.
The figure summarizes the extreme values of the bipartite negativities over the entire range of the electric-field orientation. Panels (a) and (b) show density maps of the minimum values of $N_{12}$~($=N_{23}$) and $N_{13}$, respectively, in the parameter space spanned by the electric-field magnitude and $\alpha$. The corresponding maximum values are displayed in panels (c) and (d). These density plots provide a global picture of how the electric-field strength and the degree of nonuniform spin–electric-field coupling jointly determine the achievable range of bipartite entanglement for each spin pair, thereby identifying the parameter regions where entanglement transfer is most efficient.
Although no practical mechanism is currently known for directly tuning the ratio of the KNB coupling constants, $\alpha$, our results demonstrate that the electric-field magnitude and orientation alone provide an effective means of controlling both the global and local entanglement properties of a single molecular spin system. In particular, the density plots reveal broad parameter regions in which the bipartite negativity of one spin pair reaches its maximum value, while the negativities of the remaining pairs are either completely suppressed or reduced to their minimum values. This selective localization of entanglement illustrates that an external electric field can efficiently steer quantum correlations between different spin pairs, enabling controllable redistribution of entanglement without modifying the intrinsic microscopic interactions of the molecule.

\section{Conclusion}
\label{sec5}
In this work, we investigated the simplest nontrivial molecular nanomagnet exhibiting the Katsura–Nagaosa–Balatsky (KNB) mechanism of spin-induced ferroelectricity and demonstrated its ability to realize electrically controlled bipartite entanglement transfer between different spin pairs. Although the KNB mechanism is primarily associated with magnetoelectric coupling and multiferroic phenomena~\cite{KNB1, KNB2, bro13, thakur18, XYZ, oha20, mench15, bar21, sznajd18, sznajd19, baran18, sc_KNB, JJJ, dis, oles, esa, bre1, cen19, cen21, stre20, bre2}, our study focused on its potential for manipulating quantum correlations within a molecular spin system. We showed that the spin–electric-field coupling provided by the KNB mechanism~\cite{per26} offers an efficient and versatile route for controlling both bipartite and tripartite entanglement. Depending on the electric-field strength and orientation, the entanglement can be either enhanced or suppressed, while the electric field also enables a controlled redistribution of bipartite quantum correlations among different spin pairs.
For the experimentally relevant regime in which the magnetic field remains below the saturation value and an in-plane electric field of constant magnitude rotates, we demonstrated the existence of electrically driven entanglement transfer. In this process, the bipartite negativity of one spin pair reaches its maximum, whereas the negativities of the remaining two pairs simultaneously attain their minimum values, indicating that the pairwise entanglement is effectively concentrated in a selected bond. In the highly symmetric case, characterized by homogeneous exchange interactions ($J_1=J_2$) and uniform spin–electric-field coupling ($\alpha=1$), the transfer is practically ideal: the extrema of the negativities differ from the theoretical values of $1/2$ and zero by only $10^{-8}$–$10^{-10}$.
We further showed that deviations from the symmetric case, arising from nonuniform spin–electric-field coupling ($\alpha\neq1$), unequal exchange interactions ($J_1\neq J_2$), or variations in the bond geometry, do not destroy the entanglement-transfer mechanism. Instead, these parameters provide additional degrees of freedom for tailoring the redistribution of quantum correlations and for selectively controlling the maximum entanglement attainable by individual spin pairs. Consequently, the electric-field magnitude and orientation emerge as experimentally accessible control parameters for steering both local and global entanglement properties without requiring modifications of the intrinsic microscopic interactions.

Our results demonstrate that molecular nanomagnets with KNB spin–electric coupling constitute a promising platform for electrically controlled quantum-state engineering. The ability to selectively route and localize entanglement within a single molecule may prove valuable for future implementations of molecular-scale quantum information processing, electrically driven quantum devices, and spin-based quantum technologies.

\appendix

\section{Eigenvalues and eigenvectors of the Hamiltonian}
\label{app:A}
The eigenvalues of the Hamiltonian~\eqref{Ham} are as follows:
\begin{eqnarray}\label{spec}
&&E_1 =\frac{1}{4}\left(2J_1+J_2 \right)-\frac{3}{2}B,   \nonumber\\
&&E_{2+j} =\frac{1}{4} A_j-\frac{1}{2}B,\;\; j=0,1,2  \nonumber  \\
&&E_{5+j} =\frac{1}{4} A_j+\frac{1}{2}B,\;\;\;\; j=0,1,2 \nonumber  \\
&&E_8 =\frac{1}{4}\left(2J_1+J_2 \right)+\frac{3}{2}B,
\end{eqnarray}
where the following notations for the roots of cubic equation are introduced:
\begin{eqnarray}
&& A_j=2\sqrt{Q}  \cos \left(\frac{\omega+2 \pi j}{3}\right)-\frac{2 J_1  + J_2}{3}, \nonumber\\
&& \omega=\arccos \frac{R}{\sqrt{{Q}^3}} \nonumber \\
&& R=\frac{8}{27}\left[9E^2\left(C_1\sin^2\phi+C_2\sin^2\theta\right)+C_3\right],\nonumber\\
&& Q=\frac{4}{9}\left[3E^2\left(C_4\sin^2\phi+2\sin^2\theta\right)+C_5\right],
\label{A2}
\end{eqnarray}
where the coefficients are
\begin{eqnarray}
&&C_1=6\alpha J_1 \cos\theta+\left(\cos 2\theta-\alpha^2\right)\left(J_2-J_1\right)-3J_2,\nonumber\\
&&C_2=4J_2-J_1, \nonumber \\
&&C_3=3J_1J_2\left(13J_1+2J_2\right)-10J_1^3-8J_2^3,  \\
&&C_4=2\cos 2\theta+\alpha^2, \nonumber \\
&&C_5=7J_1^2-2J_1J_2+4J_2^2.\nonumber
\end{eqnarray}
The eigenvectors have the following form:
\begin{align}\label{eq:ES}
& | \psi_1 \rangle = | \uparrow \uparrow \uparrow \rangle   \nonumber\\
& | \psi_{2+j} \rangle = \frac{|\downarrow \uparrow \uparrow \rangle + M_{j} | \uparrow \downarrow \uparrow \rangle + L_{j} |\uparrow \uparrow \downarrow \rangle}{\sqrt{1+|L_{j}|^2 + |M_{j}|^2 }} \nonumber  \\
& | \psi_{5+j} \rangle = \frac{|\uparrow\downarrow \downarrow  \rangle + M_{j} | \downarrow \uparrow \downarrow \rangle + L_{j} | \downarrow \downarrow \uparrow \rangle}{\sqrt{1+|L_{j}|^2 + |M_{j}|^2 }}  \nonumber \\
& | \psi_8 \rangle = | \downarrow \downarrow \downarrow \rangle   \nonumber \\ \nonumber
& L_{j}=\frac{( A_{j}-J_{2}) (3 J_{2}+A_{j})-4 \alpha ^2
   E^2 \sin ^2\phi}{2T_j^+} \nonumber  \\
& M_{j}= \frac{T_j^-}{T_j^+} \quad j=0,1,2\nonumber \\
& T_j^{\pm}=J_{1} (3 J_{2}+A_{j})\pm2 \alpha  E^2\sin\phi \sin (\theta \pm\phi ) \nonumber \\
   &\qquad+ i {E} [({J_{2}}+{A_{j}}) \sin (\theta \mp\phi )\mp2 \alpha  {J_{1}} \sin\phi\nonumber \\
 &\qquad+2{J_{2}} \sin (\theta \pm\phi )].
\end{align}
The corresponding eigenvalues and the eigenvectors obtained for $E=0$ are written as follows:
\begin{eqnarray}\label{eigval}
{E}_{2,5}&=&-\frac{3}{4}J_{2}\mp \frac12 B, \nonumber\\
{E}_{3,6}&=&\frac{1}{4}J_{2}-J_{1}\mp\frac 12 B, \\
{E}_{4,7}&=&\frac{1}{4}J_{2}+\frac 12 J_{1}\mp \frac 12 B,\nonumber
\end{eqnarray}
and
\begin{eqnarray}
&& | \psi_2 \rangle_{0} = \frac{1}{\sqrt{2 }} \left(|\downarrow \uparrow \uparrow  \rangle -  | \uparrow \uparrow \downarrow\rangle\right)   \nonumber\\
&& | \psi_3 \rangle_{0} = \frac{1}{\sqrt{6}} \left(|\downarrow \uparrow \uparrow  \rangle- 2|\uparrow \downarrow  \uparrow \rangle +  | \uparrow \uparrow \downarrow\rangle\right) \nonumber \\
&& | \psi_4 \rangle_{0} =\frac{1}{\sqrt{3}}\left(|\downarrow \uparrow \uparrow  \rangle + |\uparrow \downarrow  \uparrow \rangle +  | \uparrow \uparrow \downarrow\rangle\right) \nonumber \\
&& | \psi_5 \rangle_{0} = \frac{1}{\sqrt{2 }} \left(|\uparrow \downarrow\downarrow  \rangle -  | \downarrow\downarrow\uparrow \rangle\right)  \\
&& | \psi_6 \rangle_{0} = \frac{1}{\sqrt{6}} \left(| \uparrow\downarrow\downarrow  \rangle- 2|\downarrow\uparrow\downarrow   \rangle +  | \downarrow\downarrow\uparrow  \rangle \right) \nonumber  \\
&& | \psi_7 \rangle_{0} =\frac{1}{\sqrt{3}}  \left(| \uparrow\downarrow\downarrow  \rangle+|\downarrow\uparrow\downarrow   \rangle +  | \downarrow\downarrow\uparrow  \rangle \right) \nonumber
\end{eqnarray}
The eigenstates $|\psi_1\rangle$ and $|\psi_8\rangle$ with the corresponding eigenvalues are the same as in Eq.~\eqref{eq:ES}.

\section*{Acknowledgements}
A.C., V.O., and Z.A. acknowledge partial financial support form ANSEF (Grants No. PS-condmatth-2884 and PS-condmatth-3273), V.O. and Z.A. also express their gratitude to CS RA MESCS (Grants No. 21AG-1C047 and 23AA-1C032). A.G. acknowledges financial support by the Slovak Academy of Sciences through Stefan Schwarz fund No. 2024/OV1/026. A.G., J.S., and H.A.Z. were supported by Slovak Research and Development Agency under the Contract No. APVV-24-0091 and by the Grant of The Ministry of Education, Research, Development and Youth of the Slovak Republic under the Contract No. VEGA 1/0298/25.  H.A.Z. also acknowledges the financial support provided under the postdoctoral fellowship program of P J Šafárik University in Košice, Slovakia.

\bibliographystyle{apsrev4-2}
\bibliography{KNB_triangle}

\end{document}